\magnification=1200
\pretolerance=10000
\baselineskip=24pt

\centerline{\bf NON-DIFFUSIVE PROPAGATION}                              
\centerline{\bf OF ULTRA HIGH ENERGY COSMIC RAYS}
\bigskip
\centerline{G.A. Medina Tanco, E.M. de Gouveia Dal Pino and J.E.Horvath}
\bigskip
\centerline{\it Instituto Astron\^omico e Geof\'\i sico}
\centerline{\it Universidade de S\~ao Paulo}
\centerline{\it Av.M.St\'efano 4200 - 
Agua Funda (04301-904) S\~ao Paulo SP - BRASIL}

\vskip 1 true cm
{\bf Abstract}

We report the results of 3-D simulations of non-diffusive propagation of 
Ultra-High Energy Cosmic Rays (UHECR) ($E > 10^{20} \, eV$) through the 
intergalactic and extended halo media. We quantify the expected angular 
and temporal correlations between the events and the sources, 
and the temporal delay between protons and gamma-ray counterparts
with a common origin for 
both halo and extragalactic origins.  
It is shown that the proposed UHECR-supergalactic plane 
source associations require either extremely high values of the halo magnetic 
field over as much as 100 $kpc$ length scale or a very large 
correlation length for the IGM, even for the largest possible 
values of the intergalactic magnetic field. It can be stated that the 
UHECR seem to point to the sources even more strongly than previously 
believed. The simulations also show that the 
calculated time delays between UHE protons and gamma-ray 
counterparts do not match the claimed GRB-UHECR associations for either 
cosmological or extended halo distance scales.  

\vfill\eject 
                              
{\bf 1. Introduction}
\bigskip

 The  recent  detection of several 
UHECR beyond $100 \; EeV$ ($ 1 \, EeV  \, =  \, 10^{18}  \, eV$)  
poses  a
challenge  for  the  understanding  of  their  sources. First, the
extreme   energies   are  difficult   to   obtain from {\it any}
astrophysical source [1]. Second, because  of 
photomeson production due to interactions  with
the  CMBR  photons,  protons loose  substantial  amounts  of
energy $\Delta E/E \, \sim \, 10-20 \% $ per collision
along their path to the Earth, leading
to a theoretical cutoff of the primary spectrum at $E_{max} \, \simeq \, 
5 \, \times \, 10^{19} \, eV$  (Greisen-Zatsepin-Kuzmin  or  GZK  cutoff) 
in the case of an extragalactic origin.  

Possible acceleration mechanisms require either 
direct accelerators  (high  magnetic
fields  and rotation rates to produce a large induced e.m.f.
or  reconnect  the field lines) or statistical accelerators
(powerful  shocks),  each one with their  own  difficulties  for
reaching UHE [1,2,3]. In 
principle both galactic and extragalactic environments are allowed. 
Several sites have been put forward to explain the origin of these UHECR.

 Specific candidates include
magnetars (pulsars with anomalously large magnetic  fields
$B \, \geq \, 10^{14} \, G$) as the  most likely
galactic source [4]. This possibility is particularly interesting because
magnetars are invoked as the sources of Soft-Gamma Repeaters
(SGR), now firmly identified with young SN remnants [5].  If
a rotating magnetar has a surface field $B$, a radius $R$, and a
rotational frequency $\Omega$, a circuit connected between the  pole
and  the last open field line 
would see an e.m.f $\sim \, \Omega^{2} B R^{3} c$.
The corresponding maximum energy that a particle can extract
from the rotating magnetar is then

$$ E_{max} \, \sim \, 1.7 \, \times \, 10^{21}
{\bigl( {\Omega \over {10^{4} \, s^{-1}}} \bigr)}^{2} \, 
{\bigl( {B \over {10^{14} \, G}} \bigr)} \, 
{\bigl( {R \over {10 \, km}} \bigr)}^{3} \, eV  \eqno(1) $$

\noindent
for direct acceleration along open field lines [1,6,7].

 A magnetar may also provide UHECR through particle acceleration in
reconnection  regions.  If  we assume  that  magnetic  loops
expand  by  Parker  instability [8,9] on the  surface, a 
one-dimensional   analysis   of   the
reconnection  shows  that the  maximum
kinetic  energy that a particle (e.g., proton)  can  extract
from the reconnection region is

$$E_{max} \, \simeq \, 2.2 \, \times \, 10^{20} 
{\biggl( {n \over {10^{8} cm^{-3}}} \biggr)}^{-1} \, 
{\biggl( {B \over {10^{14} \, G}} \biggr)} \, eV  \eqno(2) $$

\noindent
where $n$ is the particle density at the magnetosphere. 
Another version of (coherent) large-scale acceleration
has been advocated by Colgate [10] as an alternative powerful mechanism 
that may be operative in the halo.

At even larger distance scales 
FR II radio sources and AGNs  have been also considered  [11,12]
because  they  are known to be energetic and  accelerate  at
least  $e^{-}$  to form jets, although the efficiency for UHE acceleration
(through stochastic processes) is quite uncertain, and  the
GZK cutoff certainly applies to them. Other cosmological sources for the 
UHECR have been studied by Bird el al. [13] , Yoshida et al.[14], Waxman [15] 
and Milgrom and Usov [7].
Finally, models in which topological defects accelerate the primaries may 
be involved. For example, Sivaram [16] argues that a single cosmic string may 
match the UHECR energy and flux requirements, producing up to $10^{22} \, eV$
protons.  In this sense, the detection of UHECR will serve to boost the 
theoretical study of topological defect dynamics, sites  
that can not be completely excluded at present and 
remain an interesting possibility.

An even more urgent question than the details of the process capable of 
accelerating the UHECR is the distance scale to the sources and the angular 
correlation of the events with them; i.e., another 
version of the galactic vs. extragalactic debate. Arguments for the association 
of UHECR with extragalactic sources have been presented by Rachen and Biermann 
[17], Biermann [11] and Stanev et al. [18], who contend that the events 
exhibit a correlation with the 
supergalactic plane. Wydoczyk and Wolfendale [19] have discussed the 
features of a Giant Halo Model acceleration scenario, while 
Waxman [15] and Vietri [20] , and Vietri [21] have discussed the 
energy budget and observational constraints for a cosmological and coronal 
origin respectively. 
Furthermore, Milgrom and Usov [7,22] have claimed a 
positional association with gamma-ray 
bursts preceding the two highest-energy UHECR by 5.5 and 11 months 
respectively, and a related discussion on this important point has been given 
by Vietri [21] .
We must not forget that even the very nature of the primaries is 
hotly debated. Neutrinos and gammas, for example, can not be completely 
discarded (see however [23]), although the simplest hypothesis of protons being 
the primaries is reasonable and will be adopted throughout this work.

With the aim of helping to elucidate the problem of UHECR source 
identification we perform in this work 3-D simulations of individual 
proton trajectories propagated through the IGM, halo and ISM components.
We calculate the  
angle $\alpha$ between the arrival direction and the line-of-sight to the 
source and the time delay 
$\Delta \tau$ of the UHECR with respect to photons.
The next sections describe the numerical simulations performed to 
address these points and present the results for the use of model builders.

\bigskip

\noindent
{\bf 2. Simulations}
\bigskip

 To study the quantitative features of UHECR propagation in the ISM, halo 
and IGM we
have performed 3-D numerical simulations of the trajectories of the
energetic  particles in the intergalactic  and  interstellar
magnetic  fields, including the relevant losses due  to  their
interaction with the CMBR photons.
 As a reasonable working hypothesis we assume that  
the magnetic field  inside  the
propagation  region is uniform on scales  smaller  than  its
correlation length $L_{c}$ [11,24]. 
The 3-D space is divided in domains randomly generated out of a normal 
distribution of average length-scale $L_{c}$ and $10 \, \%$ standard 
deviation.
The magnetic field is assumed to be 
homogeneous inside each domain and randomly oriented
with respect to the field in adjacent domains (c.f. [11]).
Particles are injected  into the system at different  energies
and pitch angles with respect to the line of sight $\theta$. 
Our aim is to follow explicitely the 3-D trajectories of the protons through 
the magnetized media. We note that, although the path of the 
particles is expected to be a random walk in most aspects (and thus 
somewhat predictable by simple arguments), only 3-D numerical 
calculations will allow an accurate explicit determination 
of $\alpha$ and $\Delta \tau$, an information which is necessarily lost 
in diffusion-type schemes [3,24,25]. In this way we avoid to make a dangerous 
extrapolation of the diffusion coefficients to the highest energies.

Since we are interested in the UHE regime alone, we have only considered 
 the dominant loss mechanism of 
photomeson production [24,26-28], neglecting $e^{+} e^{-}$ production due to
its very small inelasticity for $E \, \geq \, 10^{20} \, eV$. 
To model this process we
use the characteristic collision time between a proton  of
energy $E \, \gg \, m_{p} c^{2}$ and a 
photon at the CMBR equilibrium
temperature $T \, = \, 3 \, K$ 
calculated by [10].  We
considered  that in each collision the energy loss 
$\Delta E / E$ increases linearly  from
$0.13$,  at $E \, = \, 10^{20} \, eV$, to 
$0.22$, at $2.3 \, \times \, 10^{20} \, eV$ at the
$(3/2,3/2)$ resonance [27], being constant  at higher
energies (this sets a {\it lower} limit estimate to the actual losses due to
the onset of multiple pion production).
 The  output  parameters of the simulation are the 
time of flight between the injection site (the  source) and the
detector $\tau$, the arrival energy $E$ 
and deviation angle $\alpha$ from the line  of
sight for each propagated particle.

Several numerical experiments have been performed by injecting $\sim \, 8 \, 
\times \, 10^{5}$ particles having a $N(E) \, \propto \, E^{-2}$ spectrum 
in the energy interval $2 \, \times \, 10^{19} \, eV \, - \, 1 \, \times \, 
10^{23} \, eV$ , except in the calculations leading to Fig. 1 
where we have injected 
monochromatic spectra ranging from $10^{21} eV$ 
to $10^{23} eV$. Although the propagated particles are tracked down to 
energies bellow $10^{20} eV$, our analysis will focuse on $E > 10^{20} eV$.
The simulations were performed for sources of UHECR at 
galactic ($10 \, kpc$), extended halo ($100 \, kpc$) 
and nearby extragalactic ($50 \, Mpc$) distance scales (different propagation
models and physical conditions will be presented elsewhere [29]). 

Motivated by the recently proposed associations of UHECR with some 
extragalactic sources [18] and a possible GRB-UHECR connection [7]
we mainly discuss the results of the extended halo and extragalactic cases.
The results of those simulations can be appreciated in Figs.1-5, 
which depict the observable quantities we are interested in 
(see captions for details). All the presented data neglects the very small 
deflection of the local (disk) ISM which does not contribute appreciably 
to the total $\alpha$. We discuss the meaning of these results in the 
next Section. 

\bigskip

{\bf 3. Results and Discussion}
\bigskip

\noindent
From the results presented in Fig. 1, and in agreement 
with the GZK cutoff expectation, our work also shows that an 
extragalactic origin is possible if $d \, \leq \, 50 \, Mpc$, 
otherwise the injection energy must be unreasonably high. 
This is in good agreement with previous diffusion-type calculations 
(e.g. [3,23,24,28]) and serves as a check of the code against those works 
mostly interested in the energy distribution of the arrival particles.

Let us discuss first the case of extended halo sources [7,10,21]. Fig. 2a 
shows the arrival angle $\alpha$ as a function of the energy $E$ for a source 
located inside a maximally magnetized ($B_{H} \, = \, 10^{-6} \, G$) 
extended halo at a distance $d \, = \, 100 \, kpc$. As expected, this 
dependence is quite strong. However, in the energy regime we are interested 
in $\alpha$ varies from $\sim \, 10^{o}$ to $\sim \, 6^{o}$ for the highest 
energy observed events. Therefore, in the hypothesis of a common halo 
origin of GRB and UHECR, the error circles that can be determined for the 
latter are {\it less} restrictive than those of GRB, making any 
individual positional association very uncertain. Furthermore, given the 
low value of the UHECR flux $\sim \, 5 \, \times \, 10^{-12} \, erg \, 
cm^{-2} \, s^{-1} \, sr^{-1}$ it seems impossible to collect enough events 
for any other statistical positional correlation with known halo objects 
to make sense. We have checked that these results are 
robust with respect to variations in 
$B_{H}$ and $L_{c}$ as long as 
$10^{-9} \, G \, < \, B_{H} \, > \, 10^{-6} \, G$ and $L{c} \, < \, 3 \, kpc$.

As first stressed by Migrom and Usov [7] and 
other authors, the existence of a {\it temporal} 
correlation between GRB and UHECR would be very important to understand the 
location and physics of the sources. To investigate this point we have plotted 
in Fig. 2b the probability density distribution $P \, (\Delta \tau)$ of the 
arrival time delay $\Delta \tau$ between a photon and a UHE proton originated 
at the same event. Our explicit following of the trajectories shows that a 
considerable temporal spread is present and the expected value of 
$\Delta \tau$ peaks at $\sim \, 10^{3} \, yr$. A delay $\leq \, 10 \, yr$ 
has a probability of $\leq \, few \, \times \, 10^{-4}$ for protons of 
$10^{20} \, eV \, < \, E \, < \, 3 \, \times \, 10^{20} \, eV$. This means 
that if we assume a common GRB-UHECR source with $\sim \, 1 \, yr$ delay, 
the total energy in UHE protons of the the event should be $\sim \, 10^{57} \, 
erg$ for isotropic emission. The numbers quantify the brief discussion 
given in [7] and show that the association hypothesis faces severe 
problems for this distance scale. Needless to say, the problems 
of having a temporal correlation as short as $\sim \, 1 \, yr$ are 
worse for the other currently favored GRB distance scale, namely a cosmological 
one which we do not address. The result is again not sensitive to changes 
in $B_{H}$ and $L_{c}$ within the previously given range. 
Moreover, the combined 
results suggest that a correlation between GRBs and lower energy protons 
($10^{19} \, eV$ energy bump is even weaker (see for example [30] 
for a recent search).

We turn now to the extragalactic case. The basic features of an extragalactic 
source injecting protons at $d \, = \, 50 \, Mpc$ are shown in Figs. 3-5 
(see captions for details). We begin discussing the influence of the IGM 
alone and later include the combined effect of a halo for different 
assumptions. Let us consider first $B_{IGM} \, = \, 10^{-9} \, G$ and 
$L_{c} \, = \, 1 Mpc$, hereafter denoted as the fiducial case (see [31]). 
From Fig.3 (lower curve) it can be appreciated that the IGM by itself is 
not enough to produce considerable deflection for protons of 
$E \, \geq \, 10^{20} \, eV$. Therefore, within an error circle of at most 
$8^{o}$, the particles point to their sources. These error circles reduce to 
a mere $2^{o}$ for the highest energy detected event. Thus, taken at face value,
 an association with sources lying in the supergalactic plane [18] seems to be 
unlikely (although the pair of Yakutsk-Fly's Eye events could come from a 
single source). At least two assumptions (capable in principle of changing this 
conclusion) deserve further analysis. The first one is the value of 
$B_{IGM}$ which, even if it is considered a reasonable upper limit [31], 
is extracted from rotation measurements that involve $L_{c}$ and hence 
depends on it. $B_{IGM}$ is known to scale approximately as $L_{c}^{-1/2}$. 
Fig. 4 shows the effect of varying $L_{c}$ over a wide range on the arrival 
angle $\alpha$ of UHECRs. Due to energetic arguments on $B_{IGM}$, any other reasonable 
normalization allowed by the observational 
data will produce $\alpha$ values lying {\it below} 
the one in Fig. 4 and thus this curve could be considered as an upper 
limit. It seems that the deviation angle can not be increased 
by altering either $B_{IGM}$ for a given $L_{c}$ or $L_{c}$ for a 
given ({\it smaller}) $B_{IGM}$. The second major modification is the obvious 
inclusion of a halo. To maximize the deviation due to this component we have 
repeated the calculations including an extended ($R_{H} \, = \, 100 \, kpc$), 
strongly magnetized halo 
($B_{H} \, = \, 10^{-6} \, G \, ; \, L_{c} \, = \, 1.5 \, kpc$) [30], 
which is shown in the upper curve of Fig.3. In this case a source located in 
the supergalactic plane can not be excluded. We emphasize, however, 
that this is a rather extreme case, as can be checked from the curves 
presented in Fig. 5. It is seen that the deflecting power of the halo 
on extragalactic incoming particles quickly decays with its magnetic 
field $B_{H}$ and 
size $R_{H}$ (lower curve). If $B_{H}$ happens to be $0.2 \, \mu G$ as 
some observations suggest, then the halo plays almost no role at all and 
the deflection angle is entirely due to the IGM presence ($\sim \, 5^{o}$ 
in the particular case of our fiducial IGM case, Fig.5). The 
situation does not change appreciably in either case by, for example, 
doubling the halo 
correlation length (indicated by the three simulations represented by 
the symbols, see caption).

In summary we have presented a set of numerical models devised to keep the 
basic features for the propagation of protons through IGM and extended halo 
concerning the combined deviation and time delay. From the analysis of the 
results we have argued that, even if not ruled out, an association of 
UHECR ($E > 10^{20} eV$) with either GRB or supergalactic plane sources 
require rather strong assumptions on the size of the halo and/or the 
correlation length of $B_{IGM}$. It is apparent that more work on these 
subjects is needed to elucidate the final origin of UHECR. Note that at energies 
lower than $10^{20} eV$, other particle interactions besides those considered in the 
present work could be relevant and so our conclusions should not be extrapolated
to that energy range.

\noindent
{\bf Acknowledgements}

We would like to acknowledge the partial financial support of the 
CNPq and FAPESP agencies, Brazil. G.A.M.T. wishes to thank E.Ferri for 
technical support.

\noindent
{\bf References} 
\bigskip

\noindent
1)  A.M.Hillas, Ann. Rev. Astron. Astrophys. 22 (1984) 425-438.

\noindent
2) R.V.E. Lovelace, Nature 262 (1976) 649-652.

\noindent
3) R.J.Protheroe and P.A.Johnson, Astropart. Phys. 4 (1995) 253-263.

\noindent
4) V.V.Usov. Nature 357 (1992) 472-474.

\noindent
5) D.Frail and S.Kulkarni, Nature 360 (1993) 412-414.

\noindent
5) V.S.Berezinskii, in: 18th Int. Cosm. Ray Conf. (1983) 275-278.

\noindent
7) M.Milgrom and V.V.Usov, Astrophys. J. Lett. 449 (1995) L161-L164.

\noindent
8) J.G.Kirk, D.B. Melrose and E.Priest, Plasma Astrophysics 
(Springer-Verlag, Berlin 1994).

\noindent
9) P.Podsiadlowski, M.Rees and M.Ruderman, Mon. Not.R.A.S 273 (1995) 755-771.

\noindent
10) S.Colgate, in: 24th Int. Cosm. Ray Conf. (1995) 341-344.

\noindent
11) P.L.Biermann, MPI preprint (1995).

\noindent
12) J.P.Rachen, in: Proc. 17th Texas Symp. , H.Bohringer, G.E.Morfill and 
J.E.Trumper, eds. (Ann. NY Acad. Sci., NY, 1995) 468-471.

\noindent
13) D.J.Bird et al. , Astrophys.J. 441 (1995) 144-150.

\noindent
14) S.Yoshida et al., Astropart. Phys. 3 (1995) 151-159.

\noindent
15) E.Waxman, Phys. Rev. Lett. 75 (1995) 386-389.

\noindent
16) C.Sivaram, in: 24th Int. Cosm. Ray Conf. (1995) 352-355.

\noindent
17) J.P.Rachen and P.L.Biermann, Astron.Astrophys. 272 (1993) 161-167.

\noindent
18) T.Stanev, P.L.Biermann, J.Lloyd-Evans, J.P.Rachen and A.Watson, 
Phys. Rev. Lett. 75 (1995) 3056-3059.

\noindent
19) J.Wdowczyk and A.W.Wolfendale, in: 24th Int. Cosm. Ray Conf. (1995) 360-363.

\noindent
20) M.Vietri, Astrophys.J. 453 (1995) 883-889.

\noindent
21) M.Vietri, Mon. Not. R.A.S 278 (1995) L1-L4

\noindent
22) M.Milgrom and V.V.Usov, Astropart. Phys. 4 (1996) 365-369.

\noindent
23) G.Sigl, D.N.Schramm and P.Bhattacharjee, Astropart. Phys. 2 (1994) 401-411.

\noindent
24) F.A.Ahronian and J.W.Cronin, Phys.Rev.D 50 (1994) 1892-1992.

\noindent
25) A.P.Szabo and R.J.Protheroe, Astropatr. Phys. 2 (1994) 375-388.

\noindent
26) G.T.Zatsepin and V.A.Kuz'min, ZhETF Pis'ma 4 (1966) 114-116.

\noindent
27) K.Greisen, Phys. Rev. Lett. 16 (1966) 748-750.

\noindent
28) V.S.Berezinsky and S.I.Grigor'eva, Astron. Astrophys. 199 (1988) 1-12.

\noindent
29) G.A. Medina Tanco, E.M de Gouveia Dal Pino and J.E.Horvath, 
in preparation.

\noindent
30) T.Stanev, R.Schaefer and A.Watson, Astropart.Phys. 5 (1996) 75-78.

\noindent
31) P.Konberg, Rep. Prog. Phys. (1994) 325-382.

\vfill\eject

\noindent
{\bf Figure captions}

Figure 1. Arrival energy of UHE protons $E$ as a function of the distance to 
the source $d$ for several values of the (monochromatic) injection energy
$E_{\bf inj}$. 
          
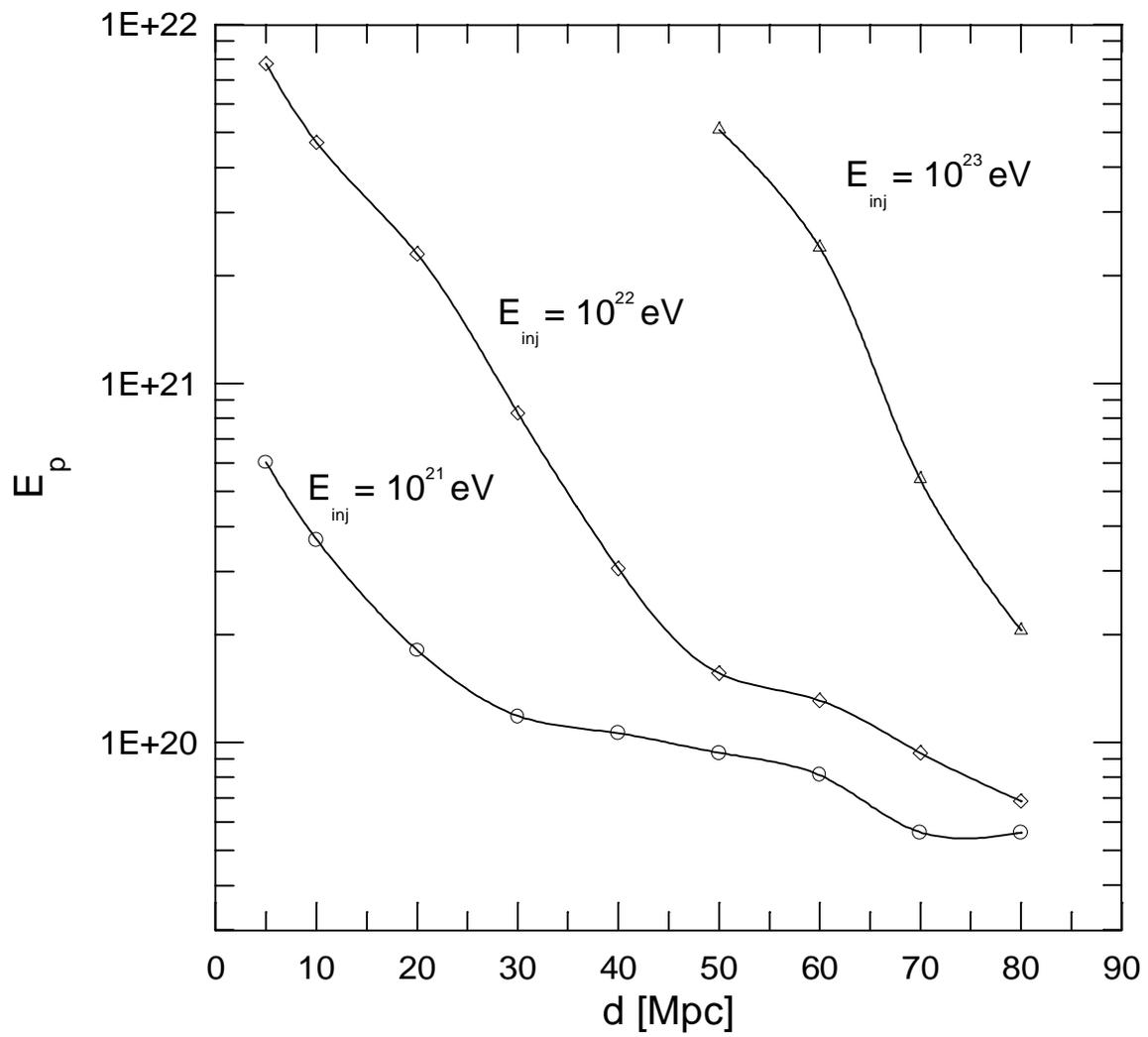
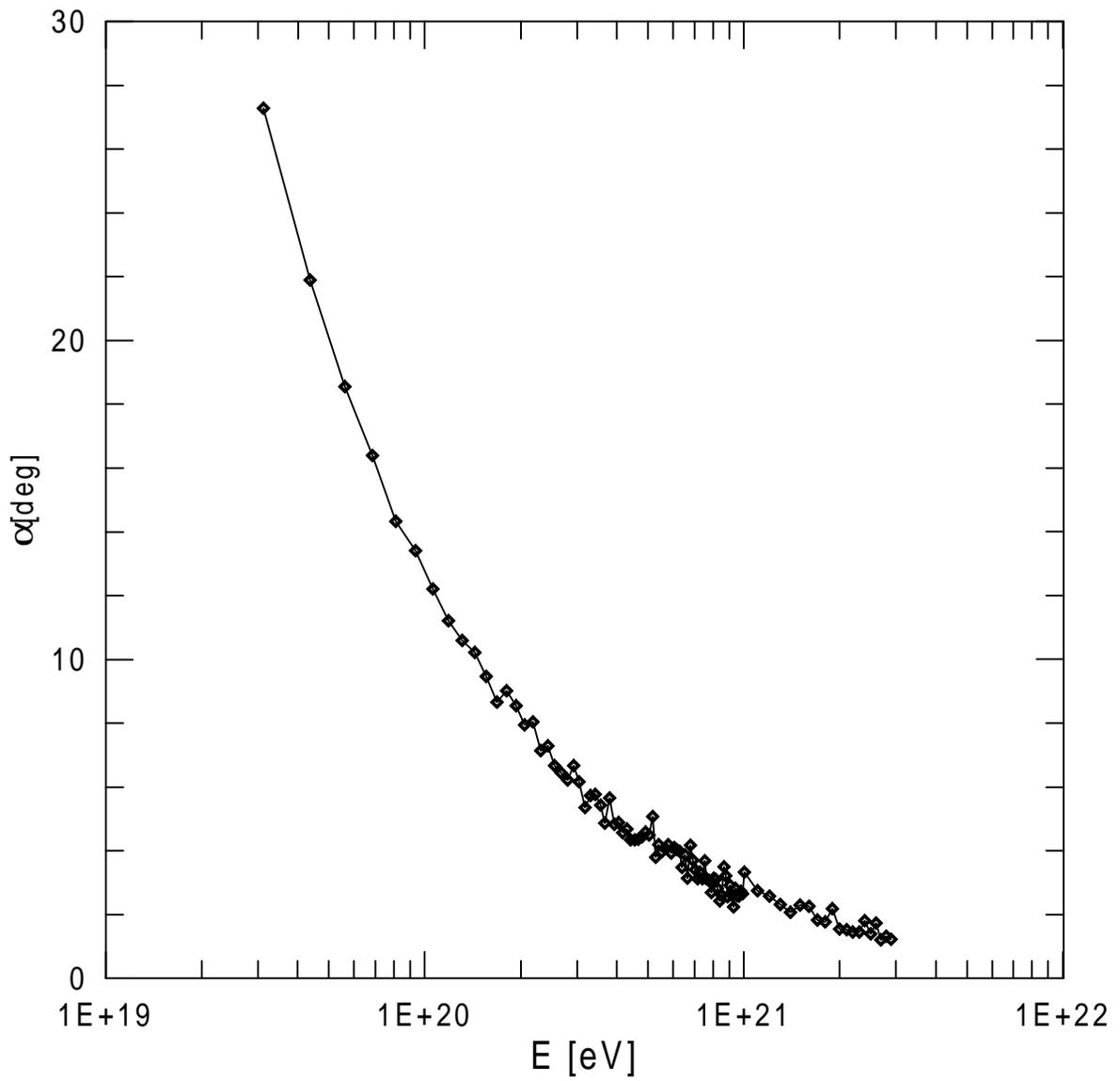
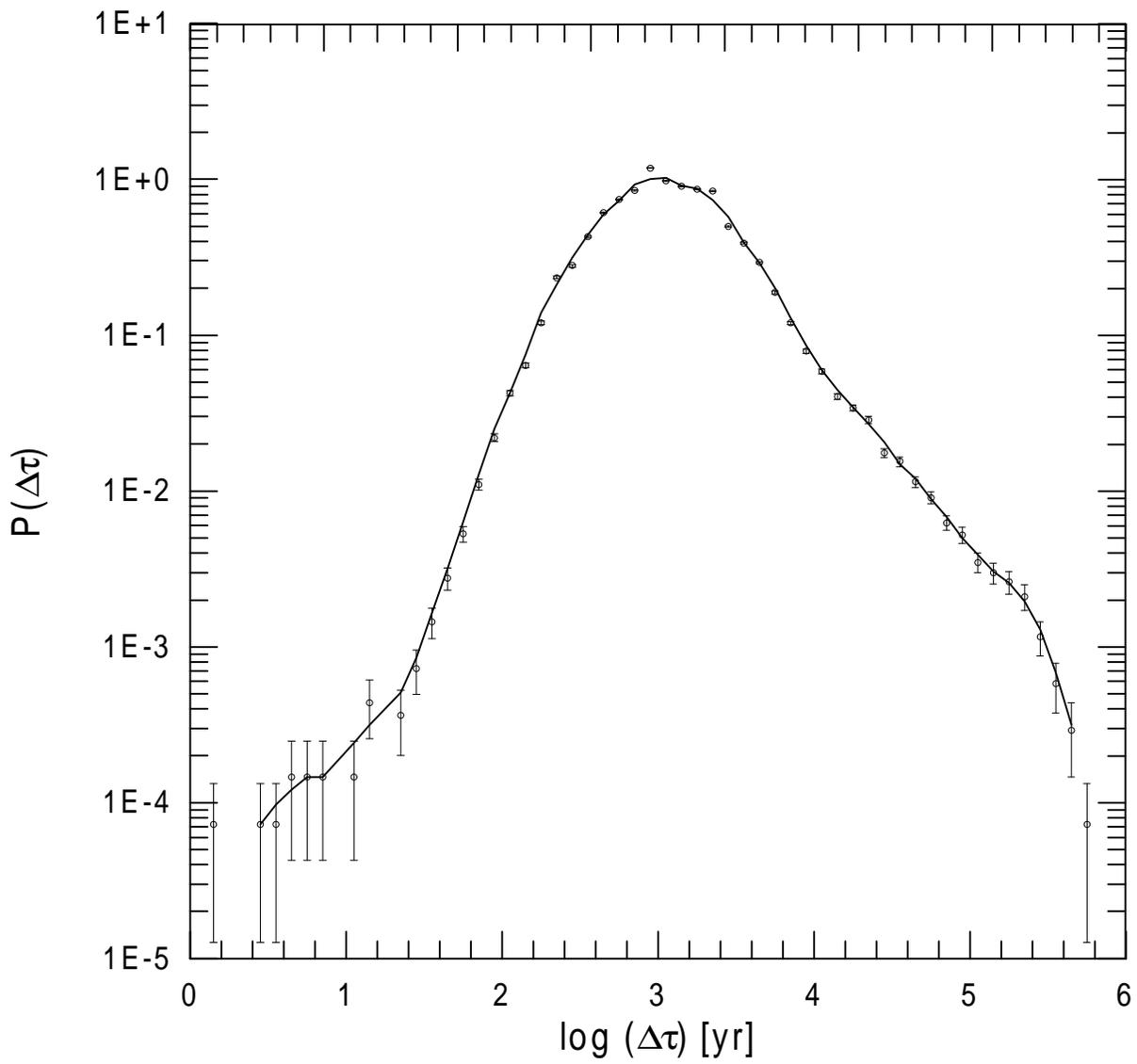
Figure 2. a) Arrival angle $\alpha$ of a proton of energy $E$. Extended halo 
case for a power-law injected spectrum (see text),source distance 
$d \, = \, 100 \, kpc$ , magnetic field strength 
$B_{H} \, = \, 10^{-6} \, G$ and $L_{c} \, = \, 1.5 \, kpc$.
2 b) Probability density distribution $P\, (\Delta \tau)$ as a function of the 
proton delay with respect to photons $\Delta \tau$, assuming simultaneous 
injection at the source. The error bars reflect the resolution of the 
simulations.

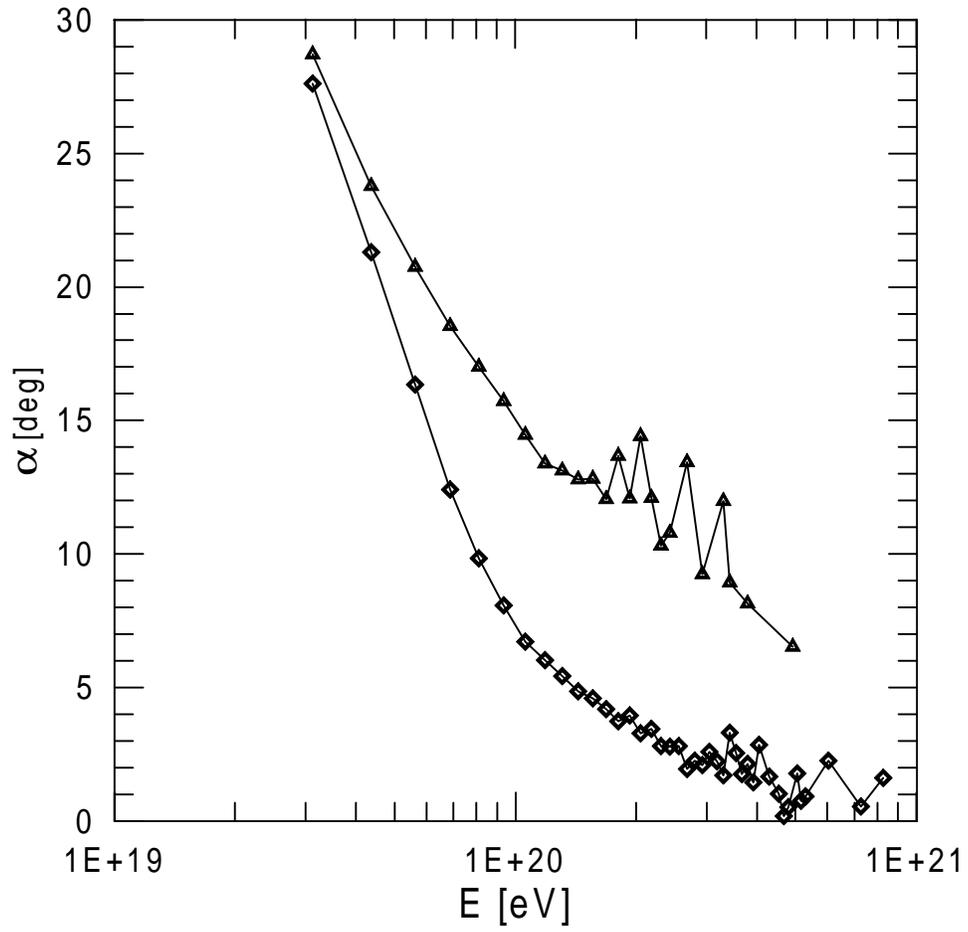
Figure 3. The same as in Fig. 2a for an extragalactic source at 
$d \, = \, 50 \, Mpc$. The curves are given for $B_{IGM} \, = \, 10^{-9} \, G$, 
$L_{c} \, = \, 1 \, Mpc$ without considering the effects of $B_{H}$ 
(lower curve) and, for the same values of the IGM, with the inclusion of 
a maximally magnetized halo having $B_{H} \, = \, 10^{-6} \, G$,  
$L_{c} \, = \, 1.5 \, kpc$ 
and size $R_{H} \, = \, 100 \, kpc$ (upper curve). See the discussion 
in the text.

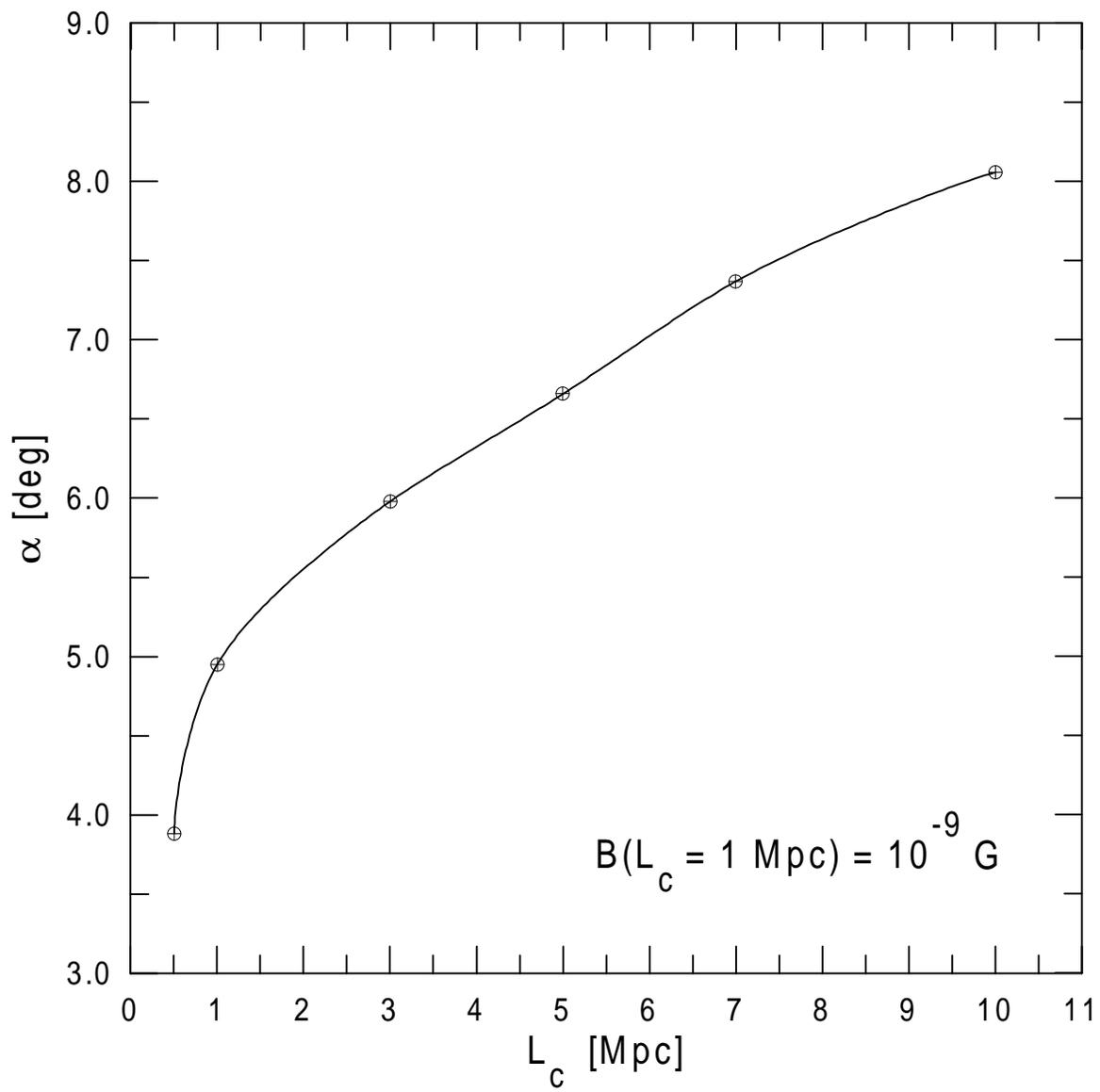
Figure 4. Upper limit to the arrival angle $\alpha$ as a 
function of $L_{c}$ allowed to increase 
from the normalization point value 
$B_{IGM} ( L_{c} \, = \, 1 \, Mpc ) \, = \, 10^{-9} \, G$. Values of 
$L_{c} \, < \, 1 \, Mpc$ would require even higher $B_{IGM}$ and begin to 
violate energetic constraints.

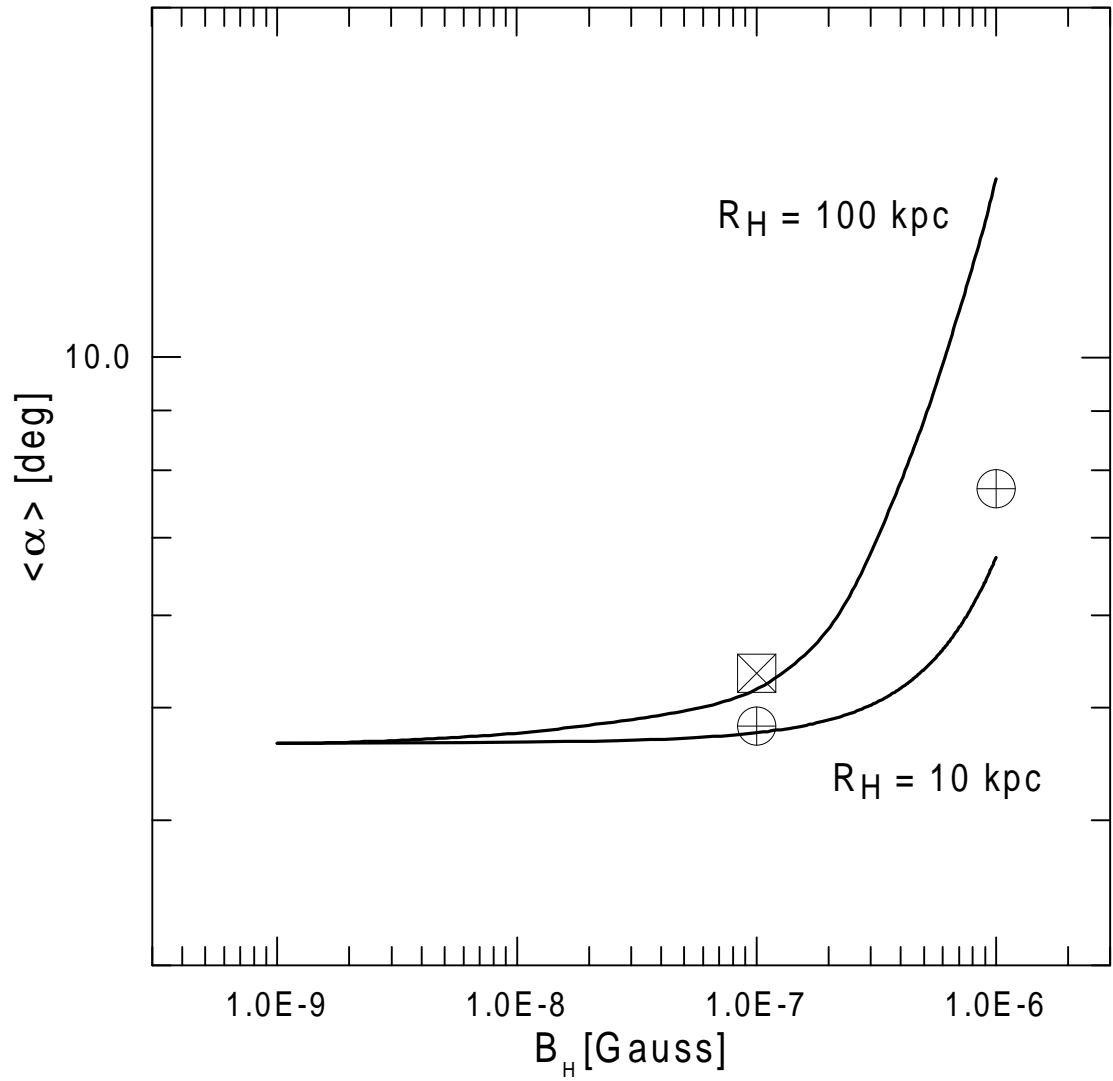
Figure 5. Influence of the halo on the arrival angle of extragalactic protons 
injected at $d \, = \, 50 \, Mpc$ with a power-law spectrum and traveling 
through an IGM having $B_{H} \, = \, 10^{-6} \, G$ ordered on 
$1 \, Mpc$ scale. Two different halo models are considered : a conservative one 
of a characteristic size $R_{H}$ of $10 \, kpc$ (lower curve) and an 
extended one with $R_{H} \, = \, 100 \, kpc$ (upper curve). The curves span all 
cases ranging from asymptotically negligible halos 
(for $B_{H} \, < \, 10^{-7} \, G$) 
to maximally deviating halos having $B_{H} \, = \, 10^{-6} \, G$ 
(end of the curves). 
(the observations actually suggest a value of about $0.2 \, \mu G$, see [31])
The curves have been calculated assuming a halo $L_{c} \, = \, 1.5 \, kpc$. 
The symbols show three simulations in which $L_{c}$ has been doubled 
for $R_{H} \, = \, 10 \, kpc$ (crossed circles) and $R_{H} \, = \, 100 \, kpc$ 
(square), showing the insensitivity of the results to this parameter.

\bigskip


\bye